\documentclass[notoc,11pt,twoside,nohyper]{JHEP} 
\usepackage{epsfig}
\usepackage{amssymb}
\newcommand\fverb{\setbox\pippobox=\hbox\bgroup\verb}
\newcommand\fverbdo{\egroup\medskip\noindent%
            \fbox{\unhbox\pippobox}\ }
\newcommand\fverbit{\egroup\item[\fbox{\unhbox\pippobox}]}
\newbox\pippobox
\title{Heavy quark production in $\gamma\gamma$ collisions revisited}

\author{by J. Ch\'{y}la\thanks{Work done in Centre for Particle Physics
supported by the Ministry of Education of the Czech Republic under the
project LN00A006}\\
    Institute of Physics, Na Slovance 2, Prague 8, Czech Republic\\
    E-mail: \email{chyla@fzu.cz}}
\received{\today}       
\accepted{\today}       
\abstract{
Heavy quark production in $\gamma\gamma$ collisions is reanalyzed.
It is argued that evaluating the cross section
$\sigma(\gamma\gamma\rightarrow Q\overline{Q})$ in a well-defined
renormalization scheme requires the inclusion of direct photon
contributions up to the order $\alpha^2\alpha_s^2$. The order
$\alpha^2\alpha_s^2$ direct photon contributions are furthermore needed
for factorization scale invariance of the sum of direct and resolved
photon contributions. The importance of quantitative analysis of
renormalization and factorization scale dependence of the approximation
currently used for the evaluation of
$\sigma(\gamma\gamma\rightarrow Q\overline{Q})$ is emphasized as the
only way of estimating the theoretical uncertainty related to the
ambiguity in choosing these scales.}

\preprint{  }
\begin{document}


\section{Introduction}
\label{sec:intro}
Heavy quark production in $\gamma\gamma$ collisions has recently
received increased theoretical attention \cite{zerwas,kramer,laenen}
motivated by new experimental data on $c\overline{c}$ and
$b\overline{b}$ production from LEP2 \cite{lepcc},
which provide particularly suitable framework for the confrontation
of perturbative QCD with data. The results obtained so far and shown,
for instance, in Fig. 2 of \cite{eric} are mixed and inconclusive.

The data on $c\overline{c}$ production coming from different experiments
are not mutually quite compatible but all fall into
the broad band of theoretical prediction. This is possible because this
band reflecting the current uncertainties of theoretical calculations is
so broad that it can accommodate results differing by a factor of 2.5!
The analogous band narrows somewhat for the $b\overline{b}$ production,
but even there
the lower and upper range of theoretical predictions differ by a factor
of 2. In this case the two data points available are a factor of 3 above
the median of theoretical predictions, but the experimental errors are
too large to draw any definite conclusions. If confirmed, it would,
however, be difficult to accommodate this result theoretically as one
expects perturbative QCD to be better applicable to $b\overline{b}$
production than to the $c\overline{c}$ one. In the words of
\cite{laenen} ``there is a serious discrepancy with bottom data''.

Theoretical ``errors'' come in part from the uncertainty in the
values of $m_c$ and $m_b$ but also from the dependence of finite order
QCD approximations on the renormalization and factorization scales
and schemes. On the experimental side, major part of the error bars
reflects the uncertainties in the extrapolation of visible cross section
into the full phase space, which can be reduced only if better
understanding of the theory is achieved. In such a situation it appears
useful and timely to reanalyze the theoretical framework currently used
for analyses of heavy quark production in $\gamma\gamma$ collisions,
with particular attention to the renormalization and
factorization scale invariance. This paper was motivated in part by
nontrivial results of the detailed analysis of these dependencies in
the case of heavy quark production in pp and $\overline{\mathrm{p}}$p
collisions performed in \cite{guido}. Using this  analysis as guidance,
I will discuss the approximation currently used for the evaluation of
$\sigma(\gamma\gamma\rightarrow Q\overline{Q})$ and point out its
shortcomings. In particular, I will argue that it does not represent
genuine next-to-leading order (NLO) QCD approximation and show that
the missing ingredient are direct photon contributions of the order
$\alpha^2\alpha_s^2$. These terms, which have not yet been calculated,
come in three classes, each playing its specific role.

The paper is organized as follows. In the next two sections basic facts
relevant for our discussion are recalled and the conventional treatment
of $\sigma(\gamma\gamma\rightarrow Q\overline{Q})$ reviewed. In Section
\ref{sec:discourse} two distinct attitudes toward the meaning and content
of the terms ``leading'' and ``next-to-leading'' orders of QCD are
introduced and compared. Direct and resolved photon contributions to
$\sigma(\gamma\gamma\rightarrow Q\overline{Q})$ are analyzed in Sections
\ref{sec:direct} and \ref{sec:resolved} respectively, followed in
Section \ref{sec:missing} by the discussion of phenomenological
consequences of the present analysis.

\section{Basic facts and formulae on the structure of the photon}
\label{sec:facts}
The factorization scale dependence of
parton distribution functions (PDF) of the photon is determined by the
system of coupled inhomogeneous evolution equations
for quark singlet and nonsinglet and gluon distribution functions
\begin{eqnarray}
\frac{{\mathrm d}\Sigma(x,M)}{{\mathrm d}\ln M^2}& =&
\delta_{\Sigma}k_q+P_{qq}\otimes \Sigma+ P_{qG}\otimes G,
\label{Sigmaevolution}
\\ \frac{{\mathrm d}G(x,M)}{{\mathrm d}\ln M^2} & =& k_G+
P_{Gq}\otimes \Sigma+ P_{GG}\otimes G, \label{Gevolution} \\
\frac{{\mathrm d}q_{\mathrm {NS}}(x,M)}{{\mathrm d}\ln M^2}& =&
\delta_{\mathrm {NS}} k_q+P_{\mathrm {NS}}\otimes q_{\mathrm{NS}},
\label{NSevolution}
\end{eqnarray}
where
$\delta_{\mathrm{NS}} \equiv 6n_f\left(\langle e^4\rangle-\langle
e^2\rangle ^2\right)$, $\delta_{\Sigma}=6n_f\langle e^2\rangle$ and
\begin{eqnarray}
k_q(x,M) & = & \frac{\alpha}{2\pi}\left[k^{(0)}_q(x)+
\frac{\alpha_s(M)}{2\pi}k_q^{(1)}(x)+
\left(\frac{\alpha_s(M)}{2\pi}\right)^2k^{(2)}_q(x)+\cdots\right],
\label{splitquark} \\ k_G(x,M) & = &
\frac{\alpha}{2\pi}\left[~~~~~~~~~~~~
\frac{\alpha_s(M)}{2\pi}k_G^{(1)}(x)+
\left(\frac{\alpha_s(M)}{2\pi}\right)^2k^{(2)}_G(x)+\cdots\;\right],
\label{splitgluon} \\ P_{ij}(x,M) & = &
~~~~~~~~~~~~~~~~~~\frac{\alpha_s(M)}{2\pi}P^{(0)}_{ij}(x) +
\left(\frac{\alpha_s(M)}{2\pi}\right)^2 P_{ij}^{(1)}(x)+\cdots.
\label{splitpij}
\end{eqnarray}
The leading order splitting functions
$k_q^{(0)}(x)=(x^2+(1-x)^2)$ and
$P^{(0)}_{ij}(x)$ are {\em unique}, while all higher order ones
$k^{(j)}_q,k^{(j)}_G,P^{(j)}_{kl},j\ge 1$ depend  on the choice of
the {\em factorization scheme} (FS). The equations
(\ref{Sigmaevolution}-\ref{NSevolution}) can be recast into evolution
equations for $q_i(x,M),\overline{q}_i(x,M)$ and $G(x,M)$ with
inhomogeneous splitting functions $k_{q_i}^{(0)}=3e_i^2k_q^{(0)}$.
The couplant $\alpha_s$ depends on the {\em renormalization scale} $\mu$
and satisfies the renormalization group equation
\begin{equation}
\frac{{\mathrm d}\alpha_s(\mu)}{{\mathrm d}\ln \mu^2}\equiv
\beta(\alpha_s(\mu))=
-\frac{\beta_0}{4\pi}\alpha_s^2(\mu)-
\frac{\beta_1}{16\pi^2}
\alpha_s^3(\mu)+\cdots,
\label{RG}
\end{equation}
where, in QCD with $n_f$ massless quark flavours, the first two
coefficients, $\beta_0=11-2n_f/3$ and $\beta_1=102-38n_f/3$, are unique,
while all the higher order ones are ambiguous. These non-unique
coefficients, together with the boundary condition on the solution of
(\ref{RG}), define the so called {\em renormalization scheme} (RS).
The boundary condition is conveniently specified by the scale parameter
$\Lambda_{\mathrm{RS}}$, which depends on the RS and corresponds to
such a value of $\mu$ for which $\alpha_s(\mu)=\infty$. The
couplant $\alpha_s$ is thus actually a function of the ratio
$\mu/\Lambda_{\mathrm{RS}}$, which behaves for large $\mu$ as
$\alpha_s(\mu/\Lambda)\approx 4\pi/\beta_0\ln(\mu^2/\Lambda^2)$.
For the sake of brevity I shall, however, drop the explicit
specification of the dependence on the RS and write $\alpha_s(\mu)$
only.

Provided perturbative expressions are summed to all orders, the results
for physical quantities are independent of both the renormalization scale
and scheme, but for any finite order approximation the
numerical results {\em do depend} on $\mu$ and RS and the consistency of
the theory merely requires that their variations with $\mu$ and RS are
of higher order of $\alpha_s$ than those taken into account. However,
as the variation of $\alpha_s(\mu)$ with $\mu$ (as well as
with the RS) is proportional to $\alpha_s^2$, we must include at least
first two consecutive nonzero powers of $\alpha_s$ in perturbative
expansions of physical quantities for the necessary cancellation to
operate and for the finite order approximation to be performed in a
well-defined RS of the couplant $\alpha_s$.
\FIGURE{\epsfig{file=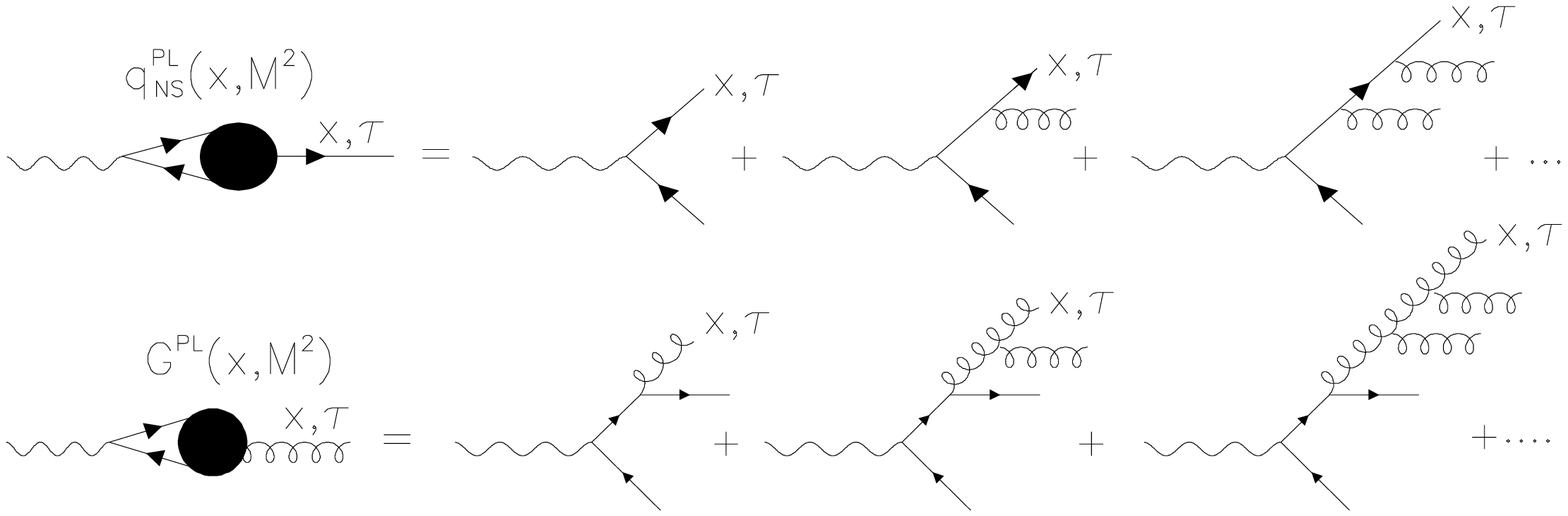,width=10cm}
\caption{
Diagrams defining the point-like parts of nonsinglet quark and
gluon distribution functions. The resummation involves integration over
parton virtualities $\tau\le M^2$ and is represented by the junction
of the blob and the $\gamma\rightarrow q\overline{q}$ vertex. Partons
going into the hard collision are denoted by $x,\tau$.}
\label{figpl}}

Due to the presence of the inhomogeneous splitting functions $k_q$ and
$k_G$ a general solution of the evolution equations
(\ref{Sigmaevolution}-\ref{NSevolution}) can be split into the
particular solutions of the full inhomogeneous equations and a general
solutions, called {\em hadron-like} (HAD), of the corresponding
homogeneous ones. A subset of the former resulting from the resummation
of contributions of diagrams in Fig. \ref{figpl}, describing multiple
parton emissions off the primary pure QED vertex
$\gamma\rightarrow q\overline{q}$ and vanishing at $M=M_0$, are called
{\em point-like} (PL) solutions. Due to the arbitrariness in the choice
of $M_0$ the separation of PDF into their point-like and hadron-like
parts is, however, ambiguous. In general we can thus write
($D=q,\overline{q},G$) \cite{smarkem}
\begin{equation}
D(x,M)= D^{\mathrm {PL}}(x,M,M_0)+D^{\mathrm{HAD}}(x,M,M_0).
\label{separation}
\end{equation}
The hadron-like parts can be represented by the same solid blobs as
those of hadrons. Appropriate graphical representation of the
point-like ones, proposed in \cite{FS,semantics}, is shown in Fig.
\ref{figpl}. By joining the primary $\gamma\rightarrow q\overline{q}$
vertex to the solid blob it reflects its perturbative origin.

The different nature of the UV renormalization of QCD quantities,
generating the renormalization scale and scheme dependence of
$\alpha_s$, and IR ``renormalization'' involved in the definition of
PDF represents the main reason for keeping the factorization and
renormalization scales $M$ and $\mu$ as independent free parameters. The
former sets the {\em upper} bound on the virtualities of partons included
in the definition of PDF and comes from IR part of loop corrections as
well as integrals over real parton emissions. The renormalization scale,
on the other hand, defines the lower bound on virtualities included in
the renormalized colour charges, masses and fields and comes entirely
from loops. There is no compelling theoretical reason why these two
scales should be identified \cite{Aur}.

\section{$\sigma(\gamma\gamma\rightarrow Q\overline{Q})$:
the conventional approach}
\label{sec:conventional}
In \cite{zerwas,kramer,laenen} the
``next--to--leading order'' QCD approximation is defined by taking the
first two terms in expansions of direct, as well as single and double
resolved photon contributions
\begin{eqnarray}
\sigma_{\mathrm{dir}} & = & \sigma_{\mathrm{dir}}^{(0)}+~~~~~
\sigma_{\mathrm{dir}}^{(1)}\alpha_s(\mu)+
\sigma_{\mathrm{dir}}^{(2)}(M,\mu)\alpha_s^2(\mu)+
\sigma_{\mathrm{dir}}^{(3)}(M,\mu)\alpha_s^3(\mu)+\cdots,\label{dir}\\
\sigma_{\mathrm{sr}} & = & ~~~~~~~~
\sigma_{\mathrm{sr}}^{(1)}(M)\alpha_s(\mu)+
\sigma_{\mathrm{sr}}^{(2)}(M,\mu)\alpha_s^2(\mu)+
\sigma_{\mathrm{sr}}^{(3)}(M,\mu)\alpha_s^3(\mu)+\cdots, \label{sr}\\
\sigma_{\mathrm{dr}} & = & ~~~~~~~~~~~~~~~~~~~~~~~~~~~~~~~~
\sigma_{\mathrm{dr}}^{(2)}(M)\alpha_s^2(\mu)+
\sigma_{\mathrm{dr}}^{(3)}(M,\mu)\alpha_s^3(\mu)+\cdots \label{dr}
\end{eqnarray}
to the total cross section
$\sigma(\gamma\gamma\rightarrow Q\overline{Q})$ for the heavy quark
pair production in $\gamma\gamma$ collisions
\begin{equation}
\sigma(\gamma\gamma\rightarrow Q\overline{Q})=
\sigma_{\mathrm{dir}}+\sigma_{\mathrm{sr}}+\sigma_{\mathrm{dr}}.
\label{total}
\end{equation}
In \cite{zerwas,kramer,laenen} three light quarks ($u,d,s$) were
considered as intrinsic for the evaluation of
$\sigma(\gamma\gamma\rightarrow c\overline{c})$ and four ($u,d,s,c$)
in the case of $b\overline{b}$ production, which is appropriate taking
into account that $m_b\gg m_c$.
Although the general strategy of the GRV group is to consider as
intrinsic quarks (of both the photon and proton) the light quarks
$u,d,s$ only, the GRV set used in \cite{zerwas} parameterizes the
effects of heavy quark production far above their thresholds by
means of the charm and bottom quark distribution functions and is
thus compatible with   \cite{zerwas}.

Note that while
$\sigma_{\mathrm{dir}}^{(0)}$ and $\sigma_{\mathrm{dir}}^{(1)}$ are
functions of $s/m_Q^2$ only, the lowest order single and double
resolved photon contributions $\sigma_{\mathrm{sr}}^{(1)}$ and
$\sigma_{\mathrm{dr}}^{(2)}$ depend, via PDF of the light quarks and
gluons, also on the factorization scale $M$, assumed for simplicity to
be the same for both colliding photons. All higher order coefficients
in (\ref{dir}-\ref{dr}) depend on both the factorization and
renormalization scales $M$ and $\mu$. The lowest order term in
(\ref{dir}) comes from pure QED and equals
\begin{equation}
\sigma^{(0)}_{\mathrm{dir}}=\sigma_0
\left[\left(1+\frac{4m_Q^2}{s}-\frac{8m_Q^4}{s^2}\right)
\ln\frac{1+\beta}{1-\beta}-\beta\left(1+\frac{4m_Q^2}{s}\right)\right],
\label{sigma0}
\end{equation}
where $s$ denotes the square of total $\gamma\gamma$ collision energy,
$\beta\equiv \sqrt{1-4m_Q^2/s}$ and
$\sigma_0\equiv 12\pi\alpha^2e_Q^4/s$. For the direct photon
contribution (\ref{dir}) the renormalization and factorization scales
$\mu$ and $M$ appear first in $\sigma_{\mathrm{dir}}^{(2)}$, which,
however, is not included in \cite{zerwas,kramer,laenen}.

Defined in this way, the direct, single resolved and double resolved
contributions start and end in (\ref{dir}-\ref{dr}) at different powers
of $\alpha_s$. In the \cite{zerwas,kramer,laenen}
this is justified by claiming that PDF of the photon behave as
$\alpha/\alpha_s$, and consequently all three expansions
(\ref{dir}-\ref{dr}) start and end at the same powers $(\alpha_s)^0=1$
and $\alpha_s$, respectively. However, as argued in detail in
\cite{factor}, the logarithm $\ln M^2$ characterizing the large $M$
behaviour of PDF of the photon cannot be interpreted as $1/\alpha_s(M)$,
as it comes from integration over
the transverse degree of freedom of the purely QED vertex
$\gamma\rightarrow q\overline{q}$! If QCD is switched off by sending,
for fixed $M_0$, $\Lambda\rightarrow 0$, quark and gluon distribution
functions of the photon approach their finite QED expressions
\begin{equation}
q_i(x,M)\rightarrow
q_i^{\mathrm{QED}}(x,M)\equiv\frac{\alpha}{2\pi}3e_i^2k_q^{(0)}(x)
\ln\frac{M^2}{M_0^2},~~~~~~
G(x,M)\rightarrow G^{\mathrm{QED}}(x,M)=0,
\label{qQED}
\end{equation}
where $M_0$ regularizes the parallel singularity
coming from the $\gamma\rightarrow q\overline{q}$ vertex
by setting the lower limit on the virtuality of
the quark or antiquark going to the hard collision. The relation
(\ref{qQED}) then implies that the single and double resolved photon
contributions (\ref{sr},\ref{dr}) vanish in this limit, because
$\alpha_s(\mu)$ does so. In the absence of QCD we thus get, as we must,
the purely QED result $\sigma_{\mathrm{dir}}^{(0)}$. Had the PDF of the
photon really behaved as $\alpha/\alpha_s$, we would, on the other hand,
expect finite contributions from the lowest order single and double
resolved photon contributions (\ref{sr},\ref{dr}) even in the limit of
switching QCD off. I will discuss this limit, together with the large
$M$ behaviour of PDF of the photon in detail in subsection
\ref{subsec:1respl}.
Taking into account the separation (\ref{separation}) we can distinguish
5 classes of resolved photon contributions:
\begin{itemize}
\item Single resolved photon using
  \begin{itemize}
    \item hadron-like parts of PDF ($\sigma_{\mathrm{srh}}$),
    \item point-like parts of PDF ($\sigma_{\mathrm{srp}}$).
  \end{itemize}
\item Double resolved photon using
  \begin{itemize}
    \item hadron-like parts of PDF on both sides
    ($\sigma_{\mathrm{drhh}}$),
    \item hadron-like parts of PDF on one side and
           point-like ones on the other ($\sigma_{\mathrm{drhp}}$),
    \item point-like parts of PDF on both sides
    ($\sigma_{\mathrm{drpp}}$).
  \end{itemize}
\end{itemize}
With this subdivision of $\sigma_{\mathrm{sr}}$ and
$\sigma_{\mathrm{dr}}$ in mind we can rewrite (\ref{total})
as follows
\begin{equation}
\sigma(\gamma\gamma\rightarrow Q\overline{Q})=
\sigma_{\mathrm{dir}}+\sigma_{\mathrm{srh}}+\sigma_{\mathrm{srp}}+
\sigma_{\mathrm{drhh}}+\sigma_{\mathrm{drhp}}+\sigma_{\mathrm{drpp}}.
\label{totalsep}
\end{equation}

\section{Discourse on semantics: defining LO and NLO in QCD}
\label{sec:discourse}
Although the definition of the concepts ``leading-order'' (LO) and
``next-to-leading-order'' (NLO) QCD approximations is primarily a
matter of semantics, it is in my view preferable to use the terminology
that respects the basic fact that we have in mind {\em orders
of QCD}, rather than the total number of terms taken into account
in expansions (\ref{dir}-\ref{dr}). Counting only the
powers of $\alpha_s$ will allow us to associate the term ``NLO''
with calculations performed in a well-defined renormalization scheme
of the couplant $\alpha_s$ as well as factorization scheme of PDF.
In this context it is useful to recall the meaning of theses terms for
three physical quantities related to (\ref{total}).

\subsection{$\sigma(\mathrm{e}^+\mathrm{e}^-
\rightarrow{\mathrm{hadrons}})$}
\label{subsec:hadrons}
For proper treatment of the direct photon contribution
$\sigma_{\mathrm{dir}}$ in (\ref{dir}), the total cross section for
e$^+$e$^-$ annihilations into hadrons at the cms energy $Q$
provides particularly suitable guidance. For $n_f$ effectively
massless quark flavors we have
\begin{equation}
\sigma_{\mathrm{had}}(Q)=
             \sigma_{\mathrm{had}}^{(0)}(Q)+
\alpha_s(\mu)\sigma_{\mathrm{had}}^{(1)}(Q)+
\alpha_s^2(\mu)\sigma_{\mathrm{had}}^{(2)}(Q/\mu)+\cdots=
\sigma_{\mathrm{had}}^{(0)}(Q)(1+r(Q)),
\label{Rlarge}
\end{equation}
where the lowest order term $\sigma_{\mathrm{had}}^{(0)}(Q)\equiv
(4\pi\alpha^2/Q^2)\sum_fe_f^2$ comes,
similarly as $\sigma_{\mathrm{dir}}^{(0)}$ in (\ref{dir}), from pure
QED, namely the diagram in Fig. \ref{eenlo}a, whereas genuine QCD
effects are contained in
\begin{equation}
r(Q)=\frac{\alpha_s(\mu)}{\pi}\left[1+\frac{\alpha_s(\mu)}
{\pi}r_1(Q/\mu)+\cdots\right].
\label{rsmall}
\end{equation}
The lowest order contribution to $r(Q)$ results from the sum of the
integral over the cross section for real gluon emissions
(Fig. \ref{eenlo}b) with the interference term between the basic QED
diagram of Fig. \ref{eenlo}a and one loop corrections to it, shown in
Fig. \ref{eenlo}c. The next-to-lowest order contribution to
(\ref{rsmall}), $(\alpha_s/\pi)^2r_1$, results from summing the integral
over the double parton emissions, exemplified by the diagrams in Fig.
\ref{eenlo}d,
with the integral over the interference term between single gluon
emission (Fig. \ref{eenlo}b) and one loop corrections to it
(for examples see Fig. \ref{eenlo}e), and the interference term between
the QED contribution of Fig. \ref{eenlo}a and the two loop corrections
to it (not shown). At each order of $\alpha_s$ the finiteness of the sum is
guaranteed by the KLN theorem.
\FIGURE{\epsfig{file=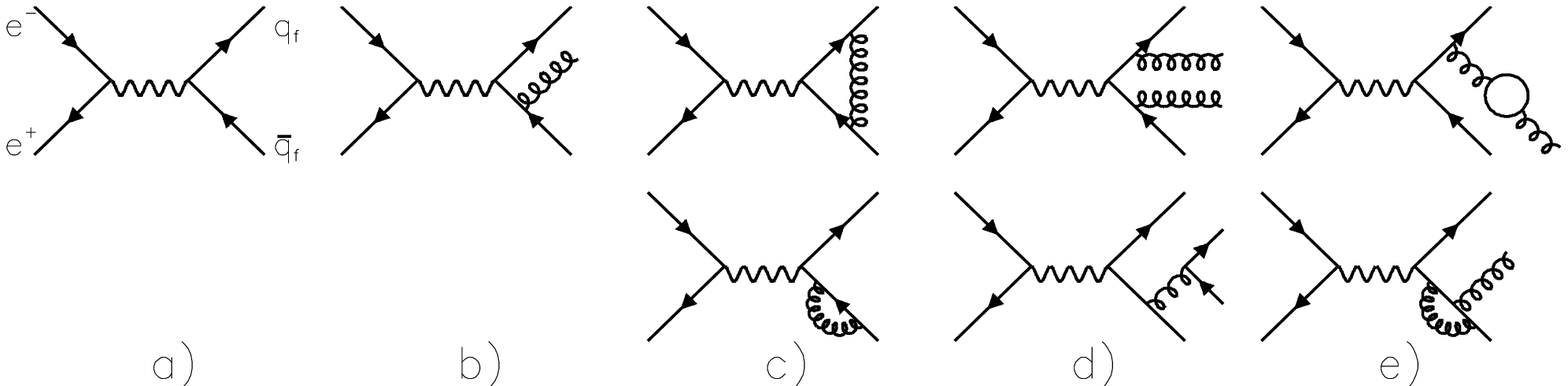,width=\textwidth}
\caption{Feynman diagrams contributing to the cross section
$\sigma_{\mathrm{ee}}(Q)$ up to the order $\alpha^2\alpha_s^2$.
Diagrams with real gluon emissions off and loop corrections on the
upper quark leg are understood to be included.}
\label{eenlo}}

For the quantity (\ref{Rlarge}) nobody calls the lowest order term
$\sigma_{\mathrm{had}}^{(0)}(Q)$ the ``leading-order'' and the next
one, i.e. $\sigma_{\mathrm{had}}^{(0)}(Q)\alpha_s(\mu)/\pi$, the
``next--to--leading order'' QCD approximations, but these terms
are applied to genuine QCD effects contained in $r(Q)$.
To work in a well-defined
renormalization scheme of $\alpha_s$ requires including in
(\ref{rsmall}) at least first two consecutive nonzero powers of
$\alpha_s(\mu)$, because only the coefficient $r_1(Q/\mu)/\pi^2$
standing by $\alpha_s^2$ depends on
the renormalization scale and scheme. The renormalization scale
invariance of the NLO approximation to $\sigma_{\mathrm{had}}(Q)$
implies that the implicit $\mu$-dependence of the leading order term
$\alpha_s(\mu)/\pi$ in (\ref{rsmall}) is cancelled to order
$\alpha_s^2$ by explicit $\mu$-dependence of the coefficient
$r_1(Q/\mu)/\pi^2$. The $\ln \mu^2$ term necessary for this
cancellation comes from loop corrections described by diagrams like
those in Fig. \ref{eenlo}e. Note that the loops in Fig. \ref{eenlo}c
contribute to the renormalization of $\alpha$, rather than of
$\alpha_s$.
Because $\alpha_s(\mu)$ is a monotonous function of $\mu$ spanning the
whole interval $(0,\infty)$, the inclusion of first two consecutive
nonzero powers of $\alpha_s$ is also a prerequisite for
the applicability of any of the scale fixing methods \cite{PMS,ECH,BLM}
available on the market. Without it, there is no preferred scale and,
consequently, the LO QCD approximations become entirely arbitrary.

For purely perturbative quantities the association of the term ``NLO
QCD approximation'' with a well-defined renormalization scheme is a
generally accepted convention, in my view worth retaining for physical
quantities in any hard scattering process.

\subsection{$F_2^{\mathrm{p}}(x,Q^2)$}
\label{subsec:F2p}
For physical quantities involving PDF proper definition of the terms
``leading'' and ``next-to-leading'' order concerns beside the
renormalization scale and scheme of $\alpha_s$ also the factorization
scale and scheme of these PDF.

For proton the structure functions $F_2^{\mathrm{p}}(x,Q^2)$ is given as
the convolution of PDF with the corresponding coefficient functions.
Dropping for simplicity quark charges we generically have
\begin{equation}
F_2^{\mathrm{p}}(x,Q^2)=q(M)\otimes C_q(Q/M)+G(M)\otimes G_G(Q/M),
\label{F2p}
\end{equation}
where quark and gluon distribution functions satisfy the homogeneous
evolution equations and
\begin{eqnarray}
C_q(x,Q/M)&=&\delta(1-x)+\frac{\alpha_s(\mu)}{2\pi}C_q^{(1)}(x,Q/M)+
\left(\frac{\alpha_s(\mu)}{2\pi}\right)^2C_q^{(2)}(x,Q/M,Q/\mu)+
\cdots\label{Cq}\\
C_G(x,Q/M)&=&0~~~~~~~~~+\frac{\alpha_s(\mu)}{2\pi}C_G^{(1)}(x,Q/M)+
\left(\frac{\alpha_s(\mu)}{2\pi}\right)^2C_G^{(2)}(x,Q/M,Q/\mu)+
\cdots.\label{CG}
\end{eqnarray}
The standard definition of LO QCD approximation to
$F_2^{\mathrm{p}}(x,Q^2)$
amounts to setting
\begin{equation}
F_{2,{\mathrm{LO}}}^{\mathrm{p}}(x,Q^2)=q(x,M),
\label{F2LO}
\end{equation}
where $M\propto Q$ and $q(x,M)$ satisfy (\ref{Sigmaevolution}-
\ref{NSevolution}) with $P_{ij}^{(0)}$ on their r.h.s. only.
At the NLO the terms proportional to $P_{ij}^{(1)}, C_q^{(1)}$ and
$C_G^{(1)}$ are included as well. All these functions depend on the
factorization scheme, and the latter two also on $M$, and must
therefore be included in any analysis which aims at working in a
well-defined such scheme. Although
$F_{2,{\mathrm{LO}}}^{\mathrm{p}}(x,Q^2)$ involves only the first,
purely QED term in (\ref{Cq}), it describes QCD effects,
because they drive the $M$-dependence of $q(x,M)$. At the NLO, the
factorization scale dependence of $q(x,M)$ in (\ref{F2p}) is cancelled
by the explicit $M$-dependence of the coefficient $C_q^{(1)}(Q/M)$,
similarly as renormalization scale dependence of $\alpha_s(\mu)$
in the first term of (\ref{rsmall}) is cancelled to the
order $\alpha_s^2$ by the $\mu$-dependence of the coefficient
$r_1(Q/\mu)$. The standard definitions of LO and NLO approximations for
$F^{\mathrm{p}}_2(x,Q^2)$ thus conform to the basic requirement that
the term ``leading order'' describes the lowest order contribution
affected by QCD effects and that working in a well-defined
renormalization as well as factorization schemes requires at least the
next-to-leading order approximation.

\subsection{$F_2^{\gamma}(x,Q^2)$}
\label{subsec:F2gamma}
The conventional analysis of photon structure function employs the same
definition of the terms ``leading'' and ``next-to-leading'' orders as
that used in \cite{zerwas,kramer,laenen}. I have discussed its
shortcomings in \cite{factor} and will therefore merely recall the
main points relevant for this paper. For the real photon
$F_2^{\gamma}(x,Q^2)$ is given as the sum of convolutions
\begin{eqnarray}
\frac{1}{x}F_2^{\gamma}(x,Q^2)&=& q_{\mathrm{NS}}(M)\otimes
C_q(Q/M)+\frac{\alpha}{2\pi}\delta_{\mathrm{NS}}C_{\gamma}+
\label{NSpart} \nonumber\\ & & \langle e^2\rangle \Sigma(M)\otimes
C_q(Q/M)+\frac{\alpha}{2\pi} \langle
e^2\rangle\delta_{\Sigma}C_{\gamma}+ \langle
e^2\rangle G(M)\otimes C_G(Q/M)
\label{S+Gpart}
\end{eqnarray}
of PDF of the photon with the coefficient functions $C_q$ and $C_G$,
given in (\ref{Cq}-\ref{CG}), and
\begin{equation}
C_{\gamma}(x,Q/M)=C_{\gamma}^{(0)}(x,Q/M)+
\frac{\alpha_s(\mu)}{2\pi}C_{\gamma}^{(1)}(x,Q/M)+
\left(\frac{\alpha_s(\mu)}{2\pi}\right)^2C_{\gamma}^{(1)}(x,Q/M)+\cdots,
\label{Cgamma}
\end{equation}
where the lowest order term in (\ref{Cgamma})
\begin{equation}
C_{\gamma}^{(0)}(x,Q/M)=\left(x^2+(1-x)^2\right)
\left[\ln\frac{M^2}{Q^2}+\ln\frac{1-x}{x}\right]+8x(1-x)-1
\label{C0}
\end{equation}
comes, similarly as $k_q^{(0)}$ in (\ref{splitquark}) and
$\sigma_{\mathrm{had}}^{(0)}$ in (\ref{Rlarge}), from pure QED.
Throughout this subsection I will discuss only the point-like part of
$F_2^{\gamma}(x,Q^2)$, which is related to the point-like parts of PDF
of the photon. The QED contribution to $F_2^{\gamma}(x,Q^2)$
\begin{equation}
\frac{1}{x}F_2^{\gamma,\mathrm{QED}}(x,Q^2)=
\sum_{i=1}^{n_f}e_i^2\left(q_i^{\mathrm{QED}}(x,M)+
\overline{q}_i^{\mathrm{QED}}(x,M)\right)+
\frac{\alpha}{2\pi}6n_f\langle e^4\rangle C_{\gamma}^{(0)}(x,Q/M)
\label{lowest}
\end{equation}
comes from the diagram in Fig. \ref{f2gamma}a and is actually
independent of
$M$ because the $M$-dependence of $q_i^{\mathrm{QED}}(x,M)$ is cancelled
by that of $C_{\gamma}^{(0)}(x,Q/M)$. Switching on QCD implies that
terms proportional to nonzero powers of $\alpha_s$ in the splitting as
well as coefficient functions (\ref{splitquark}-\ref{splitpij}) and
(\ref{Cq},\ref{CG},\ref{Cgamma}) are taken into account.
\FIGURE{\epsfig{file=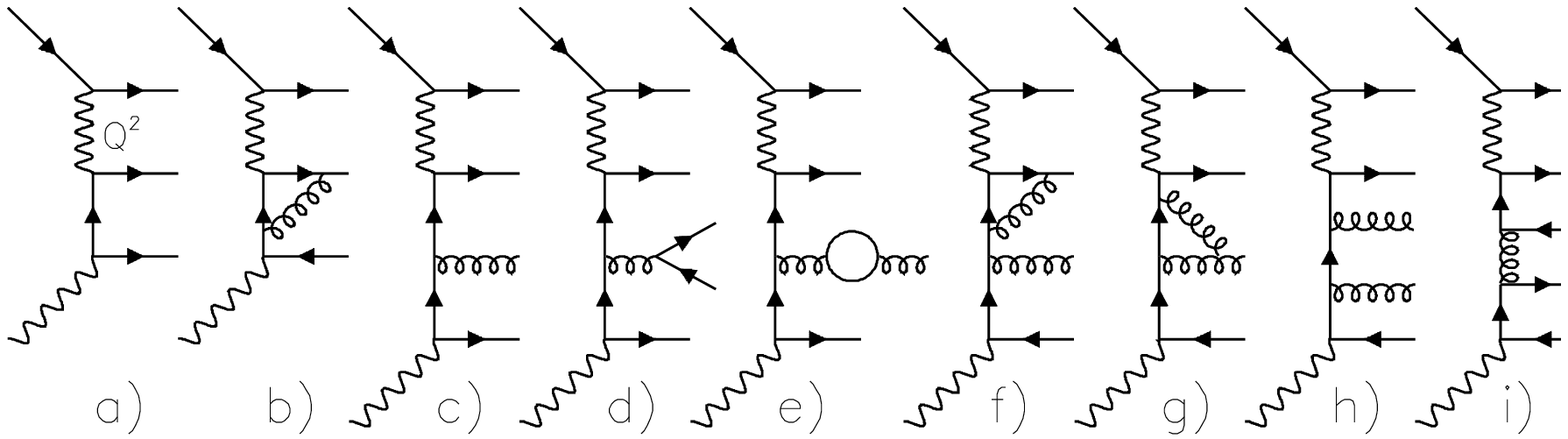,width=14cm  }
\caption{Examples of diagrams contributing to
$F_2^{\gamma}$ in QED (a), ${\cal O}(\alpha_s)$ (b-c) and
${\cal O}(\alpha_s^2)$ (d-i) QCD.}
\label{f2gamma}}

It appears natural to define as the leading order QCD approximation to
$F_2^{\gamma}$ the
one that takes into account those of the above terms proportional
to $\alpha_s$ which contribute to the ${\cal O}(\alpha_s)$ correction
to (\ref{lowest}). These are $k_q^{(1)}$ and $P_{ij}^{(0)}$ in
(\ref{splitquark}-\ref{splitpij}) and $C_q^{(1)}$ and $C_{\gamma}^{(1)}$
in (\ref{Cq}) and (\ref{Cgamma}). The remaining ${\cal O}(\alpha_s)$
terms, $k_G^{(1)}$ and $C_G^{(1)}$, do not contribute to $\alpha_s$
correction to $F_2^{\gamma}$ because $G^{\mathrm{QED}}(x,M)=0$.
The functions $k_q^{(1)},~C_q^{(1)}$ as well as
$C_{\gamma}^{(1)}$ result from evaluation of the order
$\alpha^2\alpha_s$ diagrams
\footnote{By ``order of diagram'' I mean in the case of tree diagrams
the order of its square and in the case of diagrams with a loop the
order of the interference term of this diagram with the corresponding
tree one(s).},
exemplified by those in Fig.
\ref{f2gamma}b,c. It is worth emphasizing that despite the presence of
loops involving gluons, no renormalization of $\alpha_s$ is performed
at this order, because the loops, like that in Fig. \ref{f2gamma}b,
contribute to the UV renormalization of $\alpha$, not $\alpha_s$!
In this definition the LO QCD approximation to $F_2^{\gamma}$
thus involves (see \cite{factor} for explicit formula) in addition to
the purely QED quantities $k_q^{(0)}$ and $C_{\gamma}^{(0)}$ also
$k_q^{(1)},P_{ij}^{(0)},C_q^{(1)}$ and $C_{\gamma}^{(1)}$. Note that
the conventional LO approximation takes into account out of these 6
quantities only two of them: $k_q^{(0)}$ and $P_{ij}^{(0)}$.

According to the approach advocated in \cite{factor}, the
next-to-leading order QCD approximation to $F_2^{\gamma}$
includes in addition to the quantities mentioned above also the
splitting and coefficient functions proportional to $\alpha_s^2$,
i.e. $k_q^{(2)},P_{ij}^{(1)},C_q^{(2)},C_{\gamma}^{(2)}$ and
takes into account also the contribution generated by the gluonic
splitting and coefficient functions $k_G^{(1)}$ and $C_G^{(1)}$.
The evaluation of these quantities involves order
$\alpha^2\alpha_s^2$ diagrams, exemplified by those in Fig.
\ref{f2gamma}d-i. These include loop corrections to the LO diagrams
(like those in Fig. \ref{f2gamma}e-g) as well as two parton
emissions, exemplified by diagrams in Fig. \ref{f2gamma}d,h-i.
At this order the renormalization of $\alpha_s$ starts to operate by
removing the UV divergencies coming from loops in Fig.
\ref{f2gamma}e-g (and others). Note that the conventional NLO
approximation to $F_2^{\gamma}$ (see, for instance, \cite{grv})
includes
$k_q^{(0)},k_q^{(1)},P_{ij}^{(0)},P_{ij}^{(1)},C_{\gamma}^{(0)}$ and
$C_q^{(1)}$, but not
$k_q^{(2)},C_q^{(2)},C_{\gamma}^{(1)},C_{\gamma}^{(2)}$. This is
particularly intriguing in the case of the coefficient function
$C_{\gamma}^{(1)}$ which stands in (\ref{Cgamma}) by power
$\alpha\alpha_s$ and is actually known! As
$k_q^{(2)},C_q^{(2)},C_G^{(1)}$ and $C_{\gamma}^{(2)}$,
have not yet been calculated, a complete NLO QCD approximation of
$F_2^{\gamma}$ defined in the sense advocated in \cite{factor},
cannot at the moment be constructed.

The fact that contrary to the case of $F_2^{\gamma}$, $C_q^{(1)}$
enters $F_2^{\mathrm{p}}$ first at the
NLO is due to the fact that there is no point-like, perturbatively
calculable coupling of the proton to quarks (or gluons) that would
generate the inhomogeneous splitting function analogous to $k_q^{(0)}$,
and thus the pure QED contribution analogous to (\ref{qQED}).

\section{Direct photon contribution to
$\sigma(\gamma\gamma\rightarrow Q\overline{Q})$}
\label{sec:direct}
In \cite{zerwas,kramer,laenen} the purely QED contribution
$\sigma_{\mathrm{dir}}^{(0)}$ is considered as the leading-order and
the sum
\begin{equation}
\sigma_{\mathrm{dir}}^{\mathrm{(01)}}\equiv
\sigma_{\mathrm{dir}}^{(0)}+\alpha_s(\mu)\sigma_{\mathrm{dir}}^{(1)}
\label{dirNLO}
\end{equation}
as the NLO approximation of the direct photon contribution
$\sigma_{\mathrm{dir}}$. Note that (\ref{dirNLO}) is
of the same form as the sum of first two terms in (\ref{Rlarge})
and $\alpha_s(\mu)$ is thus the only place where $\mu$ appears.
Consequently, $\sigma^{(01)}_{\mathrm{dir}}$ cannot be associated to a
well-defined renormalization scheme of $\alpha_s$ and therefore does
not deserve the label ``NLO'' even if the NLO expression for
$\alpha_s(\mu)$ is used therein. For QCD analysis of
$\sigma_{\mathrm{dir}}$ in a well-defined renormalization scheme the
incorporation of the third term in (\ref{dir}), proportional to
$\alpha^2\alpha_s^2$, is indispensable. We need in particular the
diagrams like those in Fig. \ref{direct}j-k, which involve loops
contributing to the renormalization of $\alpha_s$. The regular,
$\mu$-dependent part of their contributions provides the $\ln \mu^2$
term canceling the $\mu$ dependence of $\alpha_s(\mu)$ in the second
term of (\ref{dirNLO}).
\FIGURE{\epsfig{file=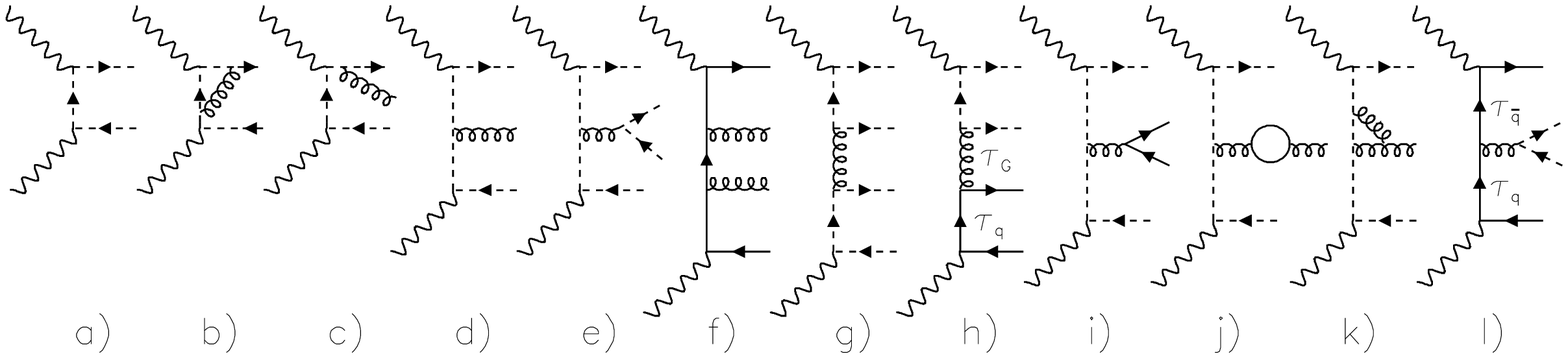,width=\textwidth}
\caption{Examples of diagrams contributing to
$\sigma_{\mathrm{dir}}$ in QED (a), leading (b-d) and next-to-leading
(e-l) orders of QCD. Heavy quark are denoted by dashed, light quarks by
solid lines.}
\label{direct}}

At the order $\alpha^2\alpha_s^2$ diagrams with light quarks appear as
well and we therefore distinguish three classes of contributions,
differing by the charge factor $CF$ ($e_Q$ and $e_q$ denoting charges of
heavy and light quarks respectively):
\begin{description}
\item{\bf Class A:} $CF=e_Q^4$. Comes from diagrams in which both
primary photons couple to heavy quarks or antiquarks. The diagrams
of this class may, as those in Fig. \ref{direct}i-j, contain also
light quarks. As emphasized above the diagrams containing loops,
like those in Fig. \ref{direct}j-k, are vital for the
renormalization of $\alpha_s$. Despite the presence of mass
singularities of individual diagrams coming from gluons and light
quarks in the final state and loops, the KLN theorem guarantees that
at each order of $\alpha_s$ the sum of all contributions of this class
to $\sigma_{\mathrm{dir}}$ is finite.
\item{\bf Class B:} $CF=e_Q^2e_q^2$.
Comes from diagrams in which one of the primary
photons couples to a heavy and the other to a light quark-antiquark
pair, like that in Fig. \ref{direct}h.
For massless light quarks this diagram has initial state mass
singularity, coming from the region of vanishing light quark virtuality
$\tau_q$, which is removed by introducing the concept of the
resolved photon. This implies subtracting from the corresponding cross
section the integral over the double pole $1/(\tau_q\tau_G)$
and putting it into the point-like part of the gluon distribution
function appearing in the lowest order single resolved photon
diagram of Fig. \ref{singlepl}a. Similarly, the single pole term
$1/\tau_q$ is subtracted and put into the light quark distribution
function entering the next-to-lowest order single resolved photon
diagram if Fig. \ref{singlepl}c.
\item{\bf Class C:} $CF=e_q^4$. Comes from diagrams in which both
photons couple to light $q\overline{q}$ pairs, as those in Fig.
\ref{direct}l. In this case the analogous subtraction procedure relates
it to the lowest order double resolved photon contribution of the
diagram in Fig. \ref{singlepl}h.
\end{description}
Because of different charge factors, the classes A, B and
C do not mix under renormalization of $\alpha_s$ and factorization of
mass singularities. Such mixing does, however, occur, within each
of these classes. At the order $\alpha^2\alpha_s^2$ all three classes
of contributions are needed for theoretical consistency, albeit each
for different reason.
The class A is needed if the calculation is to be performed in a
well-defined RS. In the next Section we shall see that classes B and C
must be taken into account for factorization scale invariance of the
sum (\ref{total}) of direct and resolved photon contributions.

\section{Resolved photon contribution to
$\sigma(\gamma\gamma\rightarrow Q\overline{Q})$}
\label{sec:resolved}
Because the factorization mechanism plays crucial role in my arguments
let me recall how it works for heavy quark production in
$\overline{\mathrm{p}}$p collisions, where the NLO approximation for
$\sigma({\overline{\mathrm{p}}\mathrm{p}}\rightarrow Q\overline{Q})$
involves convolutions of PDF with partonic cross sections up to the
order $\alpha_s^3$. Schematically
\begin{equation}
\sigma^{\mathrm{NLO}}(\overline{\mathrm{p}}\mathrm{p}
\rightarrow Q\overline{Q})=
D_1(M)\otimes\left[\alpha_s^2(\mu)\sigma^{(2)}+
\alpha_s^3(\mu)\sigma^{(3)}(M,\mu)\right]\otimes D_2(M),
\label{pp}
\end{equation}
where PDF $D_1(M)$ and $D_2(M)$ of the
colliding protons satisfy the homogeneous evolution
equations with first two terms in (\ref{splitpij}) taken into account.
Factorization scale invariance of (\ref{pp}) is guaranteed by the fact
that the $M$-dependence of $D_1(M)$ and $D_2(M)$ in the lowest order term
of (\ref{pp}) is cancelled to the order considered by the $M$-dependence
of $\sigma^{(3)}(M,\mu)$. Graphical representation of this cancellation
mechanism exploits the fact that the homogeneous part of the evolution
equations (\ref{Sigmaevolution}-\ref{NSevolution}) relate
by what I call {\em homogeneous factorization} a given diagram
with partonic cross section at the order $\alpha_s^2$ with two diagrams
(with incoming quark and gluon respectively) at the order $\alpha_s^3$
and higher. For the NLO approximation (\ref{pp}) only the lowest order
splitting functions $P_{ij}^{(0)}$ appear in this cancellation.

In \cite{guido} detailed analysis of the renormalization and
factorization
scale dependence of (\ref{pp}) has been performed under the assumption
$\mu=M$. The results, summarized in Figs. 13-15 of \cite{guido}, show
that the sensitivity of
$\sigma^{\mathrm{NLO}}(\overline{\mathrm{p}}
\mathrm{p}\rightarrow Q\overline{Q})$ to the variation of $M$ depends
sensitively on the ratio $\sqrt{s}/m_Q$, as well as on PDF of colliding
particles. They also demonstrate that contrary to the traditional
expectation the NLO QCD approximation is not necessarily less dependent
on the factorization scale $M$ than the LO ones! Whereas at $\sqrt{s}=62$
GeV $\sigma^{\mathrm{NLO}}(\overline{\mathrm{p}}\mathrm{p}
\rightarrow b\overline{b})$ is a weaker function of $M$ than
$\sigma^{\mathrm{LO}}(\overline{\mathrm{p}}\mathrm{p}
\rightarrow b\overline{b})$, and has even a stationary point,
this is not true at $\sqrt{s}=630$ GeV. There the NLO prediction is a
monotonous function of $M$, which is even steeper than the LO one!
Similar lesson follows from
comparing different processes at the same energy: whereas
$\sigma^{\mathrm{NLO}}(\overline{\mathrm{p}}\mathrm{p}
\rightarrow b\overline{b})$ has, as mentioned, a stationary point for
$\sqrt{s}=62$ GeV,
$\sigma^{\mathrm{NLO}}(\mathrm{p}\mathrm{p}\rightarrow b\overline{b})$
is again a monotonous function of $M$. To assess the stability of QCD
calculations of $\sigma(\gamma\gamma\rightarrow Q\overline{Q})$,
quantitative analysis of their factorization and renormalization scale
dependence along the lines of \cite{guido} would be very helpful,
better still with $\mu$ and $M$ treated as separate free parameters.

\subsection{Single resolved photon: the point-like part}
\label{subsec:1respl}
For the point-like part of PDF of the photon the presence of the
inhomogeneous splitting function $k_q^{(0)}$ in the evolution
equations (\ref{Sigmaevolution},\ref{NSevolution}) implies additional
relation, which I call {\em inhomogeneous factorization}, to
distinguish it from the homogeneous one introduced above, which
corresponds to $P^{(0)}_{ij}$. For instance, the lowest order
single resolved photon diagram of Fig. \ref{singlepl}a is related by
homogeneous factorization to the single resolved photon diagram of Fig.
\ref{singlepl}c. Both of these diagrams are included in the
approximation
\begin{equation}
\sigma_{\mathrm{srp}}^{(12)}(M,\mu)\equiv
\sigma_{\mathrm{srp}}^{(1)}(M)\alpha_s(\mu)+
\sigma_{\mathrm{srp}}^{(2)}(M,\mu)\alpha_s^2(\mu)
\label{srp12}
\end{equation}
employed in \cite{zerwas,kramer,laenen}. The latter diagram is related
via the inhomogeneous factorization to the direct photon diagram of Fig.
\ref{singlepl}b, which is of the order $\alpha^2\alpha_s^2$
and thus is not included in the approximation (\ref{dirNLO}) of
$\sigma_{\mathrm{dir}}$.
\FIGURE{\epsfig{file=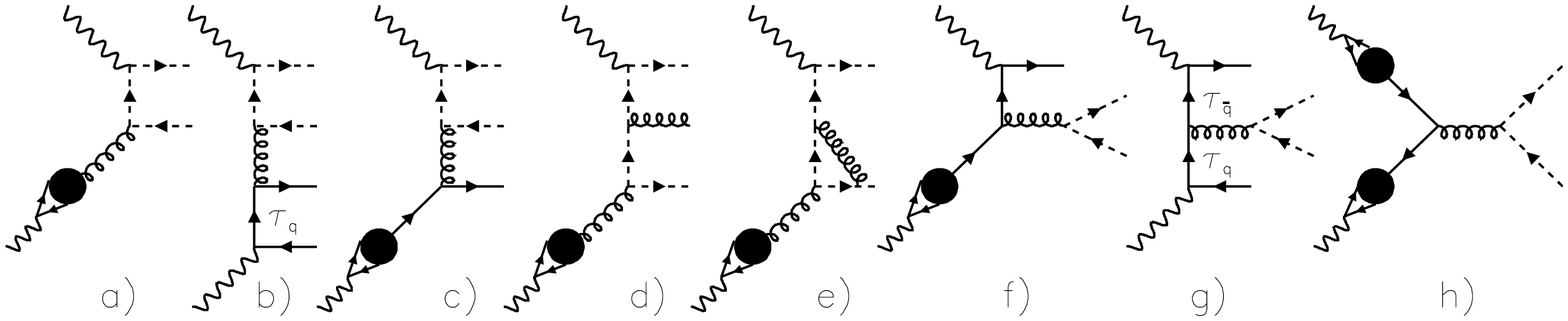,width=\textwidth}
\caption{Some of Feynman diagrams contributing to
$\sigma_{\mathrm{srp}}$ in the leading (a) and next-to-leading
(c-f) orders of QCD, together with the related direct photon diagrams
of the order $\alpha^2\alpha_s^2$ (b,g) and lowest order double
resolved photon diagram (h).}
\label{singlepl}}

The inhomogeneous factorization, which connects direct and resolved
photon diagrams at the {\em same order} of $\alpha_s$, has a simple
interpretation, reflecting its role in removing the mass singularities
of direct photon contributions associated with the point-like coupling
of initial photons to light quarks. For the direct photon diagram in
Fig. \ref{singlepl}b (which coincides with the diagram in Fig.
\ref{direct}f) this procedure has been outlined already in Section
\ref{sec:direct}. Let me recall that its finite, $M$-dependent part
contributes to class B part of
$\sigma_{\mathrm{dir}}^{(2)}$ and cancels the dominant part of the
factorization scale dependence of single resolved photon diagram in
Fig. \ref{singlepl}c. The inclusion of class B direct photon
contributions of the order $\alpha^2\alpha_s^2$ is thus vital for
factorization scale invariance of the sum (\ref{total}) of direct
and resolved photon contributions. In similar way, the factorization
scale invariance implies that single resolved photon diagram of Fig.
\ref{singlepl}f must be considered
together with class C direct photon diagram of the order
$\alpha^2\alpha_s^2$, shown in Fig. \ref{singlepl}g. This diagram is
in turn related by inhomogeneous factorization to the point-like
part of lowest order double resolved photon diagram of Fig.
\ref{singlepl}h, which is of the same order $\alpha^2\alpha_s^2$ and
will be discussed below. These considerations apparently contradict
the statement in \cite{zerwas}
\begin{quotation}
\noindent
{\em In contrast to heavy-quark production in hadron collisions we expect
small variation of the resolved-$\gamma$ cross section with $\mu$ at
the Born level, since the $\mu$ fall off of the parton cross section
$[\sim \alpha_s^{k}(\mu)]$ is neutralized asymptotically by the
increasing number of gluon and quark partons in the photons
$\sim [\alpha_s^{-1}(\mu)]^k,~k=1$ or $k=2$ for $1$- or $2$-resolved
$\gamma$.}
\end{quotation}
I will now argue that the asymptotic behaviour of Born terms in the
resolved photon contributions, mentioned in the above quotation, does
not actually imply factorization scale invariance of the
approximation used in \cite{zerwas,kramer,laenen}.
The large $M$ behaviour of the solutions of the evolution
equation (\ref{Sigmaevolution}-\ref{NSevolution}) is the same for all
three PDF and reads (in momentum space)
\begin{eqnarray}
q_{\mathrm{NS}}(n,M)&=&
\frac{\alpha}{2\pi}\delta_{\mathrm{NS}}k_q^{(0)}(n)
\left[\frac{1}{1-\overline{P}_{qq}(n)}\right]\ln M^2,\label{NSlim}\\
\Sigma(n,M)&=&\frac{\alpha}{2\pi}\delta_{\Sigma}k_q^{(0)}(n)~\left[
\frac{1-\overline{P}_{GG}(n)}{(1-\overline{P}_{qq}(n))
(1-\overline{P}_{GG}(n))-\overline{P}_{qG}(n)\overline{P}_{Gq}(n)}
\right]\ln M^2,\label{Sigmalim}\\
G(n,M)&=&\frac{\alpha}{2\pi}\delta_{\Sigma}k_q^{(0)}(n)
~\left[\frac{\overline{P}_{Gq}(n)}{(1-\overline{P}_{qq}(n))
(1-\overline{P}_{GG}(n))
-\overline{P}_{qG}(n)\overline{P}_{Gq}(n)}\right]\ln M^2,\label{Glim}\\
& = & \Sigma(n,M)\frac{\overline{P}_{Gq}(n)}{1-\overline{P}_{GG}(n)}
\label{pppp}\nonumber
\end{eqnarray}
where $\overline{P}_{ij}(n)\equiv 2P^{(0)}_{ij}(n)/\beta_0$. Although
gluons must be radiated off the primary quarks, which costs powers of
$\alpha_s$, the $\ln M^2$ rise of point-like part of quark
distribution function drives the logarithmic rise of $G(n,M)$ as well.
Consequently, $\sigma^{(1)}_{\mathrm{srp}}(M)\alpha_s(M)$, the lowest
order term in (\ref{srp12}) (where we set $\mu=M$) approaches at large
$M$ a constant, whereas in hadronic collisions the first term in
(\ref{pp}) vanishes as $M\rightarrow \infty$. This difference might
seem important, but actually is not, because the factorization scale
invariance concerns the {\em variation} of finite order approximations
with $M$, not their magnitude! Moreover, the values of $M$ used in
\cite{zerwas}, i.e. $M=\sqrt{2}m_Q$ is far from asymptotic in any case.

The relations (\ref{NSlim}-\ref{Glim}) determine the asymptotic
behaviour of PDF of the photon for {\em fixed} value of the QCD
parameter $\Lambda$. The claim that PDF of the photon behave as
$\alpha/\alpha_s(M)$ must therefore be interpreted merely as a
shorthand for this large $M$ behaviour at fixed $\Lambda$. Interpreting
$\ln M^2$ as $\ln M^2/\Lambda^2\propto \alpha_s(M)$, would lead us to
obviously
wrong conclusion that PDF of the photon blow up to infinity when
we switch QCD off by sending $\Lambda\rightarrow 0$! On the other
hand, as (\ref{NSlim}-\ref{Glim}) do not explicitly contain $\Lambda$,
they seem to hold for any value of $\Lambda$ and, consequently, one
might be tempted to conclude that they survive the limit
$\Lambda\rightarrow 0$ as well. That, however, is not the case.

To find what happens with the asymptotics (\ref{NSlim}-\ref{Glim})
when we switch QCD off we must reverse the order of limits and first
send $\Lambda\rightarrow 0$ for fixed finite $M_0$ and $M$. Doing
this we get, of course, the pure QED formulae (\ref{qQED}), which
differ dramatically for the gluon and quantitatively (i.e. by the
presence of the square brackets in (\ref{NSlim}-\ref{Sigmalim}) also
for the quark distribution functions. The corresponding
asymptotic behavior of quark and gluon distribution functions, can be
obtained formally also directly from (\ref{NSlim}-\ref{Glim}) by sending
$\beta_0\rightarrow \infty$, which implies setting
$\overline{P}_{ij}=0,\forall i,j$ there. This reflects the fact
that the factor $2/\beta_0$ multiplying in (\ref{NSlim}-\ref{Glim})
all the splitting functions $P^{(0)}_{ij}$ results from the following
limit
\begin{equation}
\lim_{M\rightarrow \infty}
\frac{\alpha_s(M)}{2\pi}\ln M^2=\frac{2}{\beta_0}
\lim_{M\rightarrow \infty}\frac{\ln M^2}{\ln (M^2/\Lambda^2)}
=\frac{2}{\beta_0}.
\label{2beta0}
\end{equation}
Sending $\alpha_s\rightarrow 0$ for fixed $M$ can be achieved either
by sending $\Lambda\rightarrow 0$ for fixed finite $M$, or, formally,
for any $M$ by sending $\beta_0\rightarrow \infty$. The latter option
is applicable even for the asymptotics (\ref{NSlim}-\ref{Glim}).
Clearly, the order of the limits $\Lambda\rightarrow 0$ and
$M\rightarrow \infty$ matters.

There is also a second difference between $Q\overline{Q}$ production
in pp and $\gamma\gamma$ collisions. At the NLO the $q^{\mathrm{PL}}$
contributes (see diagram in Fig. \ref{singlepl}c) to the second term
in (\ref{srp12}), as it does the quark distribution functions of the
proton to the second term in (\ref{pp}).
But whereas in the latter case the variation of this term
with $M$ is of one order of $\alpha_s$ higher than that taken into
account in (\ref{pp}), for $\gamma\gamma$ collisions the presence
of $k_q^{(0)}$ implies that the variation of
$\sigma_{\mathrm{srp}}^{(2)}\alpha_s^2(M)$ with $M$ is of the same
order
as this contribution itself. As argued above, the cancellation of this
dependence requires the inclusion of direct photon contributions of
class B and order $\alpha^2\alpha_s^2$.

In summary, the approximation $\sigma_{\mathrm{srp}}^{(12)}(M,\mu)$
defined in (\ref{srp12}) has the properties of genuine NLO QCD
approximation with respect to the renormalization of $\alpha_s$,
but without the inclusion of direct photon contributions of classes
B and C, like those in Fig. \ref{singlepl}b,g, it is not factorization
scale invariant to the order considered.

\subsection{Single resolved photon: the hadron-like part}
\label{subsec:1reshad}
This part of $\sigma(\gamma\gamma\rightarrow Q\overline{Q})$ has
the same structure as the direct photon contribution to the cross
section for heavy quark production in $\gamma$p collisions and I will
therefore discuss the main features of the approximation
\begin{equation}
\sigma_{\mathrm{srh}}^{(12)}(M,\mu)\equiv
\sigma_{\mathrm{srh}}^{(1)}(M)\alpha_s(\mu)+
\sigma_{\mathrm{srh}}^{(2)}(M,\mu)\alpha_s^2(\mu).
\label{srh12}
\end{equation}
in the next subsection together with those of
$\sigma_{\mathrm{drhp}}^{(23)}$, which plays the role of the resolved
photon contribution in $\gamma$p collisions.

\subsection{Double resolved photon: the hadron-like--point-like part}
\label{subsec:2reshadpl}
This part of double resolved photon contribution used in
\cite{zerwas,kramer,laenen}
\begin{equation}
\sigma_{\mathrm{drhp}}^{(23)}(M,\mu)\equiv
\sigma_{\mathrm{drhp}}^{(2)}(M)\alpha_s^2(\mu)+
\sigma_{\mathrm{drhp}}^{(3)}(M,\mu)\alpha_s^3(\mu)
\label{drhpNLO}
\end{equation}
has the same properties as the point-like part of resolved photon
contribution to $\sigma(\gamma{\mathrm{p}}\rightarrow Q\overline{Q})$.
In fact, only the first term in (\ref{drhpNLO}) needs to be included
in the sum
\begin{equation}
\sigma_{\mathrm{srh}}^{(12)}(M,\mu)+
\sigma_{\mathrm{drhp}}^{(23)}(M,\mu)
\label{reshadNLO}
\end{equation}
to make it complete NLO QCD approximation as far as both the
renormalization of $\alpha_s$ and the definition of PDF of
the photon are concerned. Note that the $M$-dependence of
$\sigma^{(2)}_{\mathrm{drhp}}(M,\mu)$ induced by the inhomogeneous
splitting function $k_q^{(0)}$ compensates the $M$-dependence of the
part of $\sigma^{(2)}_{\mathrm{srh}}(M,\mu)$ that remains after the
subtraction of the singular term coming from the vertex
$\gamma\rightarrow q\overline{q}$.

\subsection{Double resolved photon: the hadron-like--hadron-like part}
\label{subsec:2reshadhad}
This part of double resolved photon contribution has the same structure
and properties as (\ref{pp}) and, consequently, the sum of first two
terms in (\ref{dr})
\begin{equation}
\sigma_{\mathrm{drhh}}^{(23)}(M,\mu)\equiv
\sigma_{\mathrm{drhh}}^{(2)}(M)\alpha_s^2(\mu)+
\sigma_{\mathrm{drhh}}^{(3)}(M,\mu)\alpha_s^3(\mu)
\label{drhhNLO}
\end{equation}
has all the features of genuine NLO QCD approximation, both as far as
the renormalization of $\alpha_s$ and the definition of PDF are
concerned.

\subsection{Double resolved photon: the point-like--point-like part}
\label{subsec:2resplpl}
The lowest order contribution to this part of
$\sigma(\gamma\gamma\rightarrow Q\overline{Q})$ comes from the
convolution of point-like parts of PDF of both photons with $\alpha_s^2$
cross sections of $q\overline{q}\rightarrow Q\overline{Q}$ (see Fig.
\ref{singlepl}h) and $GG\rightarrow Q\overline{Q}$ (not shown)
partonic subprocesses. The former
contribution is needed to render the point-like part of single resolved
photon contribution (\ref{srp12}) coming from the diagram in Fig.
\ref{singlepl}f factorization scale invariant. But as argued in
subsection \ref{subsec:1respl}, the diagram in Fig. \ref{singlepl}h
alone is not sufficient for this purpose and class C direct photon
diagram of Fig. \ref{singlepl}g must be included as well. For the
double resolved photon contribution the same diagram is needed for
theoretical consistency already at the lowest order, i.e. for the
cancellation of part of the $M$-dependence of
$\sigma_{\mathrm{drpp}}^{(2)}(M,\mu)\alpha_s(\mu)$.

\section{Phenomenological consequences}
\label{sec:missing}
In the preceding sections I have shown that the approximation used in
\cite{zerwas,kramer,laenen}
\begin{equation}
\sigma^{\mathrm{DKF}}(M,\mu)\equiv \sigma_{\mathrm{dir}}^{(01)}(\mu)+
\sigma_{\mathrm{sr}}^{(12)}(M,\mu)+\sigma_{\mathrm{dr}}^{(23)}(M,\mu),
\label{ZKL}
\end{equation}
based on the sum of first two term in each of the expansions
(\ref{dir}-\ref{dr}), does not represent genuine NLO QCD approximation
despite the fact that it contains partonic cross sections up to order
$\alpha_s^3$ and uses the NLO form of $\alpha_s(\mu)$. The
missing direct photon contributions of the order $\alpha^2\alpha_s^2$
come in three classes, which differ by the overall charge factor,
reflecting the presence or absence of light quark pairs and the way
they couple to the primary photons. Each of these three classes is
needed for different reason, but all are needed to make the
expression
\begin{equation}
\sigma^{\mathrm{NLO}}(M,\mu)\equiv \sigma_{\mathrm{dir}}^{(02)}(M,\mu)+
\sigma_{\mathrm{sr}}^{(12)}(M,\mu)+\sigma_{\mathrm{dr}}^{(23)}(M,\mu)
\label{NLO}
\end{equation}
genuine NLO QCD approximation. Without them, (\ref{NLO}) reduces
to (\ref{ZKL}), which is complete and theoretically consistent merely
to the leading order of QCD, both as far as the renormalization of
$\alpha_s$ and the factorization of mass singularities into PDF are
concerned.

As higher order calculations involving heavy quarks are difficult to
perform, and complete calculation of $\sigma_{\mathrm{dir}}^{(2)}$
thus not in sight, it is important to find ways of estimating their
numerical importance.

\subsection{Class A}
\label{subsec:A}
This class of direct photon contributions, related to the
renormalization scale and scheme dependence of $\sigma_{\mathrm{dir}}$,
does not mix with the resolved photon contributions at any order and
can therefore be considered separately. Because the approximation
(\ref{dirNLO}) is a linear function
of $\alpha_s(\mu)$, one can by choosing appropriate
value of $\mu$ get arbitrary value of $\alpha_s(\mu)$ and thus also of
(\ref{dirNLO}), even if the NLO form of the $\mu$-dependence of
$\alpha_s(\mu)$ is used. The existence of a
well-defined ``natural'' scale, like $m_Q$ in our case,
is of no help in this respect, as $\alpha_s(\mu)$ depends beside the
renormalization scale $\mu$ also on the renormalization scheme RS, and
at the LO there is no criteria for selecting the ``natural'' RS.
Without the inclusion of the direct photon contributions of this class
and order $\alpha^2\alpha_s^2$, there is no way how to estimate the
importance of higher corrections.
According to \cite{zerwas}, $\sigma_{\mathrm{dir}}^{(1)}\alpha_s(\mu)$
gives, setting $\mu=M=\sqrt{2}m_Q$, about $25-35$\% of the pure QED
contribution $\sigma_{\mathrm{dir}}^{(0)}$. This is a sizable
correction, which can easily be doubled by inclusion of higher order
corrections of this class.

\subsection{Classes B and C: controlling the scale ambiguities}
\label{subsec:B}
These classes are required to render the sum of point-like parts of
single and double resolved photon contributions factorization scale
invariant. As none of them is included in \cite{zerwas,kramer,laenen},
the sum of resolved photon contributions
\begin{equation}
\sigma_{\mathrm{res}}^{\mathrm{DKF}}(M,\mu)
\equiv \sigma_{\mathrm{sr}}^{(12)}(M,\mu)+
\sigma_{\mathrm{dr}}^{(23)}(M,\mu)
\label{sum}
\end{equation}
used in \cite{zerwas,kramer,laenen} is expected to exhibit
monotonous dependence on the factorization scale $M$. On the other
hand, as the point-like parts of PDF of the photon can be considered
as a way of approximate evaluation of higher order perturbative
corrections \cite{smarkem}, one may wonder whether by clever choice
of the factorization scale $M$ one could approximately include direct
photon contributions of the classes B and C in the point-like parts of
single and double resolved photon ones. To establish whether this
can be arranged and how, requires, however, a quantitative analysis of
factorization and renormalization scale dependence of the expression
(\ref{sum}). Specifically, one should plot for given $s_{\gamma\gamma}$
and $m_Q$, the quantity (\ref{sum}) as a two-dimensional function of
$\mu$ and $M$. Such a plot would give us a clear idea of the stability
of (\ref{sum}) with respect to the variations of $\mu$ and $M$ and,
possibly, suggest better choices of these scales than those adopted in
\cite{zerwas} ($\mu=M=\sqrt{2}m_Q$) or \cite{kramer,laenen}
($M=2\mu=2m_T$).

\section{Summary and conclusions}
\label{sec:summary}
The fact that the conventional NLO analyses of heavy quark production
in $\gamma\gamma$ collisions do not include direct photon contributions
of the order $\alpha^2\alpha_s^2$ represents a serious shortcoming
preventing us from drawing definite conclusions from the comparison of
existing data on $c\overline{c}$ and $b\overline{b}$ production with
currently available QCD calculations.

The missing direct photon
contributions of the order $\alpha^2\alpha_s^2$ come in three classes,
depending on the overall charge factor. One of them, proportional to
$e_Q^4$, concerns exclusively the direct photon contribution to
$\sigma(\gamma\gamma\rightarrow Q\overline{Q})$, and is vital for
establishing the genuine NLO character of $\sigma_{\mathrm{dir}}$.
The other two classes of direct photon contributions of the order
$\alpha^2\alpha_s^2$ are needed for the factorization scale invariance
of the sum of direct and resolved photon contributions. To assess their
numerical relevance, quantitative investigation of the factorization
and renormalization scale dependence of the approximation (\ref{sum})
would be very helpful.

\vspace*{0.5cm}
I am grateful to M. Kr\"{a}mer for correspondence concerning the
treatment of intrinsic charm in the $b\overline{b}$ production.

\end{document}